\documentclass[letterpaper, 10 pt, journal, twoside]{ieeetran}

\usepackage{cite}	     	
\usepackage{textcomp}   	
\usepackage{lipsum}	     	
\usepackage{soul, color}   	
\usepackage[svgnames]{xcolor} 
\usepackage[utf8]{inputenc} 
\usepackage[T1]{fontenc}    
\usepackage{enumitem}  		
\usepackage{url}

\usepackage{amsmath}    	
\usepackage{amssymb}		
\usepackage{amsfonts}		
\usepackage{amsthm}    	    
\usepackage{mathtools}		
\usepackage{newtxmath} 	    

\usepackage{bm}             

\usepackage[caption=false,font=footnotesize]{subfig}
\usepackage{graphicx}		
\usepackage{booktabs}   	
\usepackage{siunitx}    
\usepackage[export]{adjustbox}  
\usepackage{floatrow}		
\usepackage{subfig}			

\usepackage{geometry}
\usepackage{hyperref}		

\graphicspath{{images/}}  
\setlist{noitemsep}       
\hypersetup{			  
    colorlinks,
    linkcolor={MediumBlue},
    citecolor={MediumBlue},
    urlcolor={MediumBlue}
}

\newtheoremstyle{mythmstyle}  
{1ex}
{1ex}
{\itshape}
{}
{\bfseries}
{. }
{ }
{\thmname{#1}\thmnumber{ #2}\thmnote{ (#3)}}  

\newtheoremstyle{myremark} 
{}
{}
{}
{}
{\bfseries}
{.}
{.5em}
{}

\theoremstyle{mythmstyle}			
\newtheorem{assumption}{Assumption}

\theoremstyle{mythmstyle}			
\newtheorem{theorem}{Theorem}

\theoremstyle{mythmstyle}			
\newtheorem{corollary}{Corollary}

\theoremstyle{mythmstyle}			

\theoremstyle{mythmstyle}			
\newtheorem{definition}{Definition}

\theoremstyle{mythmstyle}			

\theoremstyle{myremark} 

\newcommand{\T}{^{\mathrm{T}}}							
\newcommand{\B}[1]{\if#1\relax\bm{#1}\else\mathbf{#1}\fi} 
\newcommand{\R}[1]{\mathrm{#1}}						      
\newcommand{\C}[1]{\mathcal{#1}}
\newcommand{\BB}[1]{\mathbb{#1}}

\begin{document}
\newgeometry{top=60pt, left=48pt, right=48pt, bottom=45pt}

\pagestyle{empty}



\title{Synchronization of\\Networks of Piecewise-Smooth Systems}


%
\author{\IEEEauthorblockN{
Marco Coraggio\IEEEauthorrefmark{1},
Pietro DeLellis\IEEEauthorrefmark{1},
S. John Hogan\IEEEauthorrefmark{2}, and
Mario di Bernardo\IEEEauthorrefmark{1}\IEEEauthorrefmark{2}}\\
\IEEEauthorblockA{\IEEEauthorrefmark{1}\normalsize Dept. of Information Technology and Electrical Eng., Univ. of Naples Federico II, Via Claudio 21, 80125 Naples, Italy
}\\
\IEEEauthorblockA{\IEEEauthorrefmark{2}Department of Engineering Mathematics, University of Bristol, Bristol, BS8 1UB, UK\\
Email: marco.coraggio@unina.it, pietro.delellis@unina.it, s.j.hogan@bristol.ac.uk,
mario.dibernardo@unina.it}
}



\maketitle

\thispagestyle{empty} 


\begin{abstract}
We study convergence in networks of piecewise-smooth systems that commonly arise in applications to model dynamical systems whose evolution is affected by macroscopic events such as switches and impacts.
Existing approaches were typically oriented towards guaranteeing global bounded synchronizability, local stability of the synchronization manifold, or achieving synchronization by exerting a control action on each node.
Here we start by generalising existing results on QUAD systems to the case of piecewise-smooth systems, accounting for a large variety of nonlinear coupling laws.
Then, we propose that a discontinuous coupling can be used to guarantee global synchronizability of a network of $N$ piecewise-smooth agents under mild assumptions on the individual dynamics.
We provide extensive numerical simulations to gain insights on larger networks.
\end{abstract}


%


\section{Introduction}
\label{sec:Introduction}

\IEEEPARstart{W}{hen} applications are considered, it is not uncommon to find systems and devices that are described by \emph{piecewise-smooth} (\emph{PWS}) or impulsive models, such as electronic switching circuits, mechanisms affected by dry friction, firing neurons, and so on \cite{dibernardo2008piecewise, cortes2008discontinuous, liberzon2012switching}. 
If two or more of these systems are interconnected, a complex network \cite{wang2003complex, arenas2008synchronization, scardovi2009synchronization, lu2016complex} of PWS agents needs to be studied. 
A challenging open problem is to analyse the emergence of spontaneous synchronous behaviour in this class of networks. 
For example, in \cite{delellis2015convergence}, networks of non-identical PWS systems with linear diffusive coupling are studied, and a condition on the coupling gain is given such that the synchronization error is asymptotically bounded.
In \cite{coombes2016synchrony}, an extension of the \emph{Master Stability Function} (\emph{MSF}) approach to networks of PWS oscillators is presented, under some restrictive assumptions, obtaining a condition on the coupling gain to ensure local stability of the synchronous solution.
Similarly, the MSF method is applied to dry friction oscillators in \cite{marszal2016synchronization,marszal2017parameter}.
Furthermore, sufficient conditions were found in \cite{yang2013finite} for controlling coupled PWS chaotic systems towards a desired solution, provided that a discontinuous control action is added to every node in the network.
Other relevant references include \cite{danca2002synchronization, polynikis2009synchronizability, liu2012new, liu2012filippov, yang2013exponential, liu2011dissipativity}.
However, conditions cannot be found in the existing literature that guarantee global asymptotic synchronization of a network of PWS systems in the absence of an external control acting on all of the nodes.

\textit{Main contributions.\ } We begin by generalising existing results on the global convergence of QUAD systems \cite{porfiri2008criteria,delellis2011on}
adopting a mathematical framework suitable for PWS systems, through the use of the Filippov formalism. Specifically, we allow for a large variety of coupling laws, including linear diffusion where the inner coupling matrix is not positive definite.
After that, we introduce a discontinuous coupling protocol to guarantee synchronizability for a wider class of PWS systems, finding critical values of the coupling gains analytically for the case of two coupled agents of arbitrary dimension.
For the case of larger networks, we propose deployment of a discontinuous action with a multiplex network structure \cite{lombana2016multiplex}---i.e.~a network with different layers of coupling, each having its own topology.

The rest of the paper is outlined as follows.
Section \ref{sec:network_model} contains the problem statement; Section \ref{sec:mathematical_preliminaries} the mathematical preliminaries; in Section \ref{sec:convergence_analysis} theoretical results are presented concerning networks of PWS systems; then, Section \ref{sec:multiplex_networks} describes a multiplex control approach for networks of PWS agents, while conclusions are drawn in Section \ref{sec:conclusion}.

\section{Network model}
\label{sec:network_model}

We consider networks of $N$ PWS systems \cite{dibernardo2008piecewise} $\dot{\B{x}}_i = \B{f}(\B{x}_i;t)$, $i = 1, \dots, N$, where $\B{x}_i \in \mathbb{R}^n$ is the \emph{state} vector of the $i$-th agent, $t \in \BB{R}^+$ is time, and the vector field $\B{f} : \BB{R}^n \times \BB{R}^+ \rightarrow \BB{R}^n$ can be discontinuous with respect to $\B{x}_i$.
When such systems are coupled through an undirected unweighted \emph{graph} $\mathcal{G}$, they form a \emph{complex PWS network} of the form
\begin{equation}\label{eq:general_network}
\dot{\B{x}}_i = 
\B{f}(\B{x}_i; t) - \sum\nolimits_{j=1}^{N} L_{ij} \B{g}(\B{x}_i, \B{x}_j; t), \quad i = 1, \dots, N,
\end{equation}
where $L_{ij}$ is the element $(i,j)$ of the symmetric \emph{Laplacian matrix} $\B{L} \in \BB{R}^{N \times N}$ \cite{arenas2008synchronization}, and $\B{g} : \mathbb{R}^{n} \times \mathbb{R}^{n} \times \mathbb{R}^+ \rightarrow \mathbb{R}^n$ is a \emph{coupling function}.
In addition, we define $\B{x} \triangleq \begin{bmatrix} \B{x}_1\T & \cdots & \B{x}_N\T \end{bmatrix}\T \in \BB{R}^{nN}$ to be the \emph{stack of the states} of the nodes.

\begin{definition}[Synchronization]
Network \eqref{eq:general_network} achieves \emph{(complete) synchronization} if 
\begin{equation*}
\lim_{t \rightarrow + \infty} \left\Vert \B{x}_i(t) - \B{x}_j(t) \right\Vert = 0,
\quad i,j = 1, \dots, N, \ i \ne j.
\end{equation*}
\end{definition}
A network is said to be \emph{synchronizable} in the set $\Omega \subseteq \mathbb{R}^{nN}$ if synchronization is achieved for any initial condition $\B{x}(t\!=\!0) \in \Omega$; it is \emph{globally} synchronizable if $\Omega = \mathbb{R}^{nN}$.
Finally, we define the following: 
$\bar{\B{x}} \triangleq \sum_{i=1}^{N} \B{x}_i/N \in \BB{R}^n$ is the \emph{average} of the states of the nodes;
$\B{e}_i \triangleq \B{x}_i - \bar{\B{x}} \in \mathbb{R}^n$, with $i = 1, \dots, N$, are the \emph{synchronization errors};
$\B{e} \triangleq \begin{bmatrix} \B{e}_1\T & \cdots & \B{e}_N\T \end{bmatrix}\T \in \BB{R}^{nN}$ is the \emph{stack of the errors};
$e_\R{s} \triangleq \sum_{i=1}^{N} \left\lVert \B{e}_i \right\rVert_2 / N \in \BB{R}$ is the \emph{global synchronization error}, used as a metric of synchronization in the numerical examples for the sake of comparison with the theoretical estimates.

\textit{Notation.\ } 
$\C{B}_\delta(\B{z})$ is an open ball centred in $\B{z}$ with radius $\delta>0$;
$\C{S}({\BB{R}^n})$ is a collection of subsets in $\BB{R}^n$;
$\mu_{\R{L}}(\cdot)$ is the Lebesgue measure of a set; $\overline{\R{co}}(\cdot)$ is the convex closure of a set;
$\C{N}$ indicates any set with null Lebesgue measure;
$\left\lVert \cdot \right\rVert$ is the Euclidean norm;
$\left\lvert \cdot \right\rvert$ is the absolute value;
$\R{sym}(\cdot)$ is the symmetric part of a matrix;
$\R{diag}(\B{a})$ is the diagonal matrix having the elements of vector $\B{a}$ on its diagonal;
$\lambda_i(\cdot)$ is the $i$-th eigenvalue of a matrix, with the eigenvalues being sorted in an increasing fashion if they are all real (thus $\lambda_{\R{min}}(\cdot) \triangleq \lambda_1(\cdot)$);
$\B{I}_n$ is the $n \times n$ identity matrix;
$\B{0}$ is the null vector;
the expression $\B{A} > 0$ means that the matrix $\B{A}$ is positive definite (analogously for semi- and negative definiteness); $\otimes$ is the Kronecker product.


\section{Mathematical preliminaries}
\label{sec:mathematical_preliminaries}

In this section we give a series of preliminary concepts that will be later employed in Section \ref{sec:convergence_analysis}.
A condition that is widely used in the field of complex networks to characterize agents' internal dynamics is the so-called \emph{QUAD} condition \cite{delellis2011on, delellis2015convergence}.
\begin{definition}[QUADness]\label{def:quad}
    A function $\B{f} : \BB{R}^n \times \BB{R}^+ \rightarrow \BB{R}^n$ is \emph{QUAD($\B{P}$, $\B{Q}$)} if, $\forall \B{\xi}_1, \B{\xi}_2 \in \BB{R}^n$, $t \in \BB{R}^+$, $\exists \B{P}, \B{Q} \in \BB{R}^{n \times n}$ such that
    \begin{equation*}
    \left( \B{\xi}_1 - \B{\xi}_2 \right)\T \B{P} \left[ \B{f}(\B{\xi}_1; t) - \B{f}(\B{\xi}_2; t) \right] \le
    \left( \B{\xi}_1 - \B{\xi}_2 \right)\T \B{Q} \left( \B{\xi}_1 - \B{\xi}_2 \right).
    \end{equation*}
\end{definition}

\begin{assumption}\label{ass:pseudo_QUAD_coupling}
    The coupling function $\B{g} : \mathbb{R}^{n} \times \mathbb{R}^n \times \BB{R}^+ \rightarrow \mathbb{R}^n$ in \eqref{eq:general_network} is such that, $\forall \B{\xi}_1, \B{\xi}_2 \in \BB{R}^n$ and $\forall t \in \BB{R}^+$, (i) $\B{g}(\B{\xi}_1, \B{\xi}_1; t) = \B{0}$, (ii) it is antisymmetric with respect to its first two arguments, i.e.~$\B{g}(\B{\xi}_1, \B{\xi}_2; t) = - \B{g}(\B{\xi}_2, \B{\xi}_1; t)$, and (iii)
\begin{equation*}
\left( \B{\xi}_2 - \B{\xi}_1 \right)\T \B{P} \B{g}(\B{\xi}_1, \B{\xi}_2; t) \ge
\left( \B{\xi}_1 - \B{\xi}_2 \right)\T c \B{G} \left( \B{\xi}_1 - \B{\xi}_2 \right),
\end{equation*}
for some $\B{P}, \B{G} = \B{G}\T \in \BB{R}^{n \times n}$, $c \ge 0$.
\end{assumption}
Clearly, in the case of linear diffusive coupling, we have
\begin{equation}\label{eq:linear_diffusive_coupling}
\B{g}(\B{x}_i, \B{x}_j; t) = c\B{\Gamma}(\B{x}_j - \B{x}_i),
\end{equation}
with $\B{\Gamma} \in \BB{R}^{n \times n}$ and $\B{G} = \R{sym} (\B{P} \B{\Gamma})$.
The next assumption, also found in \cite{yang2013finite}, is a relaxation of QUADness, fulfilled by a wider range of piecewise-smooth dynamics.
\begin{assumption}\label{ass:pseudo_QUAD}
    The function $\B{f} : \BB{R}^n \times \BB{R}^+ \rightarrow \BB{R}^n$ in \eqref{eq:general_network} is such that,  $\forall \B{\xi}_1, \B{\xi}_2 \in \BB{R}^n$ and $\forall t \in \BB{R}^+$,
    \begin{equation*}
    \begin{split}
    \left( \B{\xi}_1 - \B{\xi}_2 \right)\T \B{P} \left[ \B{f}(\B{\xi}_1; t) - \B{f}(\B{\xi}_2; t) \right] \le
    &\left( \B{\xi}_1 - \B{\xi}_2 \right)\T \B{Q} \left( \B{\xi}_1 - \B{\xi}_2 \right) \\
    &+ \B{m}\T \left\lvert \B{\xi}_1 - \B{\xi}_2 \right\vert,
    \end{split}
    \end{equation*}
for some $\B{P}, \B{Q} \in \BB{R}^{n \times n}$, $\B{m} \in \BB{R}^n$.
\end{assumption}
%
Next, in Definitions \ref{def:filippov_set_valued_function}-\ref{def:set_valued_lie_derivative}, we briefly recall the main concepts introduced by Filippov to characterise solutions of PWS systems \cite{filipov1988differential, cortes2008discontinuous}.
In the rest of this section, let $\B{z} \in \BB{R}^n$, $t \in \BB{R}^+$, $\B{f} : \BB{R}^n \times \BB{R}^+ \rightarrow \BB{R}^n$ be a not necessarily continuous vector field, and $V : \BB{R}^n \rightarrow \BB{R}$ be a locally Lipschitz function, which is differentiable everywhere but in a zero-measure set $\Omega_V$.

\begin{definition}[Filippov set-valued function]\label{def:filippov_set_valued_function}
The \emph{Filippov set-valued function} associated to $\B{f}$ is $\C{F}[\B{f}] : \BB{R}^n \times \BB{R}^+ \rightarrow \C{S}(\BB{R}^n)$, and is given by
\begin{equation*}
\C{F}[\B{f}](\B{z}; t) \triangleq \bigcap_{\delta > 0} \bigcap_{\mu_{\R{L}}(\C{N}) = 0} \overline{\R{co}} \left\lbrace \B{f}(\C{B}_\delta(\B{z}) \ \backslash \ \C{N}; t )\right\rbrace.
\end{equation*} 
\end{definition}
Note that if $\B{f}$ is continuous, then $\C{F}[\B{f}] = \B{f}$.

\begin{definition}[Filippov solution]\label{def:filippov_solution}
A \emph{Filippov solution} is an absolutely continuous curve $\B{z}(t) : \BB{R}^+ \rightarrow \BB{R}^n$ satisfying, for almost all $t \in \BB{R}^+$, the differential inclusion
$\dot{\B{z}} \in \C{F}[f](\B{z}; t)$.
\end{definition}

\begin{definition}[Generalised gradient]\label{def:generalised_gradient}
The \emph{generalised gradient} of $V$ is $\partial V : \BB{R}^n \rightarrow \C{S}(\BB{R}^n)$, and is given by 
\begin{equation*}
\partial V(\B{z}) \triangleq \left\lbrace \lim_{k \rightarrow \infty} \frac{\partial}{\partial \B{z}} V(\B{z}_k) : \B{z}_k \rightarrow \B{z}, \ \B{z}_k \not\in \C{N} \cup \Omega_V \right\rbrace.
\end{equation*}
\end{definition}

\begin{definition}[Set-valued Lie derivative]\label{def:set_valued_lie_derivative}
The \emph{set-valued Lie derivative} $\C{L}_{\C{F}[\B{f}]} : \BB{R}^n \rightarrow \C{S}(\BB{R})$ of $V$ with respect to $\C{F}[\B{f}]$ is
\begin{equation*}
\C{L}_{\C{F}[\B{f}]} V(\B{z}) \triangleq \left\lbrace \ell \in \BB{R} : \exists \B{a} \in \C{F}[\B{f}](\B{z};t) \Rightarrow \B{v}\T \B{a} = \ell \ \forall \B{v} \in \partial V(\B{z})\right\rbrace.
\end{equation*}
\end{definition}

\section{Convergence analysis}
\label{sec:convergence_analysis}

\begin{figure}[t]
    \centering
    \includegraphics[max width=\columnwidth]{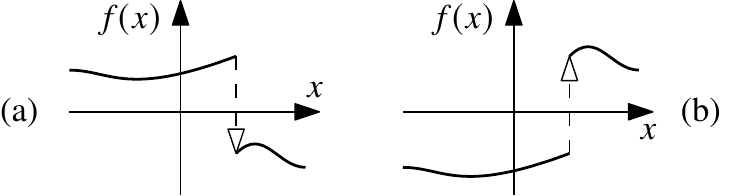}
    \caption{(a) A QUAD function; (b) a non-QUAD function.}
    \label{fig:quad_nonquad}
\end{figure}

Firstly, in Theorems \ref{thm:QUAD_Gamma_positive_definite} and \ref{thm:QUAD_Gamma_semipositive_definite}, we provide criteria to assess global synchronizability, applicable to the case that the internal agent dynamics $\B{f}$ is a QUAD function.
A certain number of discontinuous functions fall into this category, e.g.~Coulomb friction, some of relay functions, continuous but not differentiable functions like the characteristics of nonlinear resistors, scalar systems where the discontinuity causes a decrease in the value of the scalar field as the state increases (see Figure \ref{fig:quad_nonquad}), and more.
Secondly, when the individual discontinuous dynamics fails to satisfy the QUAD condition, we exploit Assumption \ref{ass:pseudo_QUAD} to investigate convergence in the case of two coupled $n$-dimensional nodes.

\subsection{Nonsmooth QUAD systems}

Next, we extend results in \cite{delellis2015convergence, porfiri2008criteria, delellis2011on} and give conditions for the global complete synchronization of PWS agents whose dynamics is QUAD, accounting for a generic nonlinear coupling function. 
Namely, considering Assumption \ref{ass:pseudo_QUAD_coupling}, Theorem \ref{thm:QUAD_Gamma_positive_definite} can be used when $\B{G} > 0$ (resp. $\B{\Gamma} > 0$ if \eqref{eq:linear_diffusive_coupling} holds), whereas Theorem \ref{thm:QUAD_Gamma_semipositive_definite} is to be employed when no assumptions on the definiteness of $\B{G}$ (resp. $\B{\Gamma}$) can be made.
Note that 
\begin{multline}\label{eq:e_i_dot}
\dot{\B{e}}_i \triangleq \B{\phi}_i(\B{x}_1, \dots, \B{x}_N; t) = \dot{\B{x}}_i - \dot{\bar{\B{x}}} =
\B{f}(\B{x}_i; t) - \sum\nolimits_{j=1}^{N}  L_{ij} \B{g}(\B{x}_i, \B{x}_j; t) \\-  \frac{1}{N} \sum\nolimits_{i=1}^{N} \left[ \B{f}(\B{x}_i; t) - \sum\nolimits_{j=1}^{N} L_{ij} \B{g}(\B{x}_i, \B{x}_j; t) \right].
\end{multline}

\begin{theorem}\label{thm:QUAD_Gamma_positive_definite}
Consider \eqref{eq:general_network} and assume that there exist $\B{P}, \B{Q}, \B{G} \in \BB{R}^{n \times n}$, $c \ge 0$, with $\B{P}, \B{G} > 0$, such that
\begin{itemize}
    \item $\B{f}$ is QUAD($\B{P}$, $\B{Q}$);
    \item $\B{g}$ verifies Assumption \ref{ass:pseudo_QUAD_coupling} with $\B{P}$ and $\B{G}$.
\end{itemize}
Then, the network is globally synchronizable if
\begin{equation}
c > c^* \triangleq \frac{\lVert \B{Q} \rVert} { \lambda_{2} (\B{L}) \lambda_{\R{min}} (\B{G}) }.
\end{equation}    
\end{theorem}
\begin{proof}  
Consider the candidate set-valued Lyapunov function $V(\B{e}) \triangleq \frac{1}{2} \sum_{i=1}^N \B{e}_i\T \B{P} \B{e}_i$.
The fact that $\B{f}$ is not continuous causes $V$ to not be differentiable.
However, employing Filippov formalism we can state that $\dot{V}(\B{x}) \in \C{V}$, where $\C{V} \triangleq \tfrac{1}{2}\sum\nolimits_{i=1}^{N} \C{L}_{\C{F}[\B{\phi}_i]} \left( \B{e}_i\T \B{P} \B{e}_i \right)$.\footnote{The \emph{sum rule} \cite{cortes2008discontinuous} was used to apply the set-valued Lie derivative operator separately to each addend in $V$.}
Hence, if $v < 0, \forall v \in \C{V}$, then $V \rightarrow 0$ and the network is globally synchronizable.
Note that, in \eqref{eq:e_i_dot}, the facts that $\B{L}$ is symmetric and $\B{g}$ is antisymmetric (w.r.t.~$\B{x}_i$ and $\B{x}_j$) imply that $\sum_{i=1}^{N} \sum_{j=1}^{N} \left[ L_{ij} \B{g}(\B{x}_i, \B{x}_j; t) \right] = 0$.
Then, from Definition \ref{def:set_valued_lie_derivative} and \eqref{eq:e_i_dot} we can write 
\begin{equation*}
\begin{split}
\C{V} = &\sum\nolimits_{i=1}^N \B{e}_i\T \B{P} \left[ \C{F}[\B{f}(\B{x}_i; t)] - \C{F}\left[ \sum\nolimits_{i=1}^{N} \frac{\B{f}(\B{x}_i; t)}{N} \right] \right] \\
&- \sum\nolimits_{i=1}^N \sum\nolimits_{j=1}^N L_{ij} \B{e}_i\T \B{P} \C{F}\left[ \B{g}(\B{x}_i, \B{x}_j; t) \right].
\end{split}
\end{equation*}
As $\sum_{i=1}^{N} \B{e}_i = 0$, we have $\sum_{i=1}^{N} \B{e}_i\T \B{P} \C{F}\left[ \sum_{i=1}^{N} \B{f}(\B{x}_i; t) / N \right] = 0$ and $\sum_{i=1}^{N} \B{e}_i\T \B{P} \C{F}[\B{f}(\bar{\B{x}}; t)] = 0$.
Thus, we can rewrite
\begin{equation*}
\begin{split}
\C{V} = &\sum\nolimits_{i=1}^N \B{e}_i\T \B{P} \left[ \C{F}[\B{f}(\B{x}_i;t)] - \C{F}\left[ \B{f}(\bar{\B{x}};t) \right] \right] \\
&- \sum\nolimits_{i=1}^N \sum\nolimits_{j=1}^N L_{ij} \B{e}_i\T \B{P} \C{F} \left[\B{g}(\B{x}_i, \B{x}_j; t)\right].
\end{split}
\end{equation*}
Focusing on a generic element $v \in \C{V}$ and exploiting the hypotheses on $\B{f}$ and $\B{g}$, we get\footnote{Recalling that $\B{L} = \B{L}\T$, and using (i), (ii), (iii) in Assumption \ref{ass:pseudo_QUAD_coupling}, we get
$- \sum\nolimits_{i=1}^N \sum\nolimits_{j=1}^N L_{ij} \B{e}_i \T \B{P} \B{g}(\B{x}_i, \B{x}_j; t) 
= - \sum\nolimits_{i=1}^N \sum\nolimits_{j>i}^N L_{ij} (\B{x}_i - \B{x}_j)\T \B{P} \B{g}(\B{x}_i, \B{x}_j; t) 
\le - c \sum\nolimits_{i=1}^N \sum\nolimits_{j>i}^N L_{ij} (\B{x}_i - \B{x}_j)\T \B{G} (\B{x}_j - \B{x}_i)
= - c \sum\nolimits_{i=1}^N \sum\nolimits_{j=1}^N L_{ij} 
\B{e}_i\T \B{G} \B{e}_j$.}
$
v \le \sum\nolimits_{i=1}^N \B{e}_i\T \B{Q} \B{e}_i  
-c \sum\nolimits_{i=1}^N \sum\nolimits_{j=1}^N L_{ij} \B{e}_i\T \B{G} \B{e}_j.
$
This inequality can be rewritten in terms of the stack of the errors $\B{e}$ as
\begin{align}
v & \le \B{e}\T \left( \B{I}_N \otimes \B{Q} - c \B{L} \otimes \B{G} \right) \B{e} \le \B{e}\T \left( \lVert \B{Q} \rVert \B{I}_N \otimes \B{I}_n - c \B{L} \otimes \B{G} \right) \B{e}\nonumber\\
&= \left\lVert \B{e} \right\rVert^2 \lVert \B{Q} \rVert - \B{e}\T \left( c \B{L} \otimes \B{G} \right) \B{e}.\label{eq:proof_branching_point_QUAD}
\end{align}
Since $\sum_{i=1}^{N} \B{e}_i = \B{0} \Leftrightarrow \sum_{i=0}^{N-1} (\B{e})_{(i-1)n+h} = 0 \ \forall h = 1, \dots, n$, we can apply Corollary 13.4.2 in \cite{godsil2013algebraic} and get\footnote{If \eqref{eq:linear_diffusive_coupling} holds,  with $\B{\Gamma}$ being an M-matrix \cite{delellis2018partial}, then a diagonal matrix $\B{M}$ exists (that is $\B{P}$) such that $\R{sym} (\B{M}\B{\Gamma}) = \R{sym} (\B{G}) = \B{G} > 0$.}
$
v \le \left\lVert \B{e} \right\rVert^2 \left[ \lVert \B{Q} \rVert - c \lambda_{2} (\B{L}) \lambda_{\R{min}} (\B{G}) \right].
$
Therefore, if $c > c^*$, $\dot{V}(\B{e}) < - \alpha \left\lVert \B{e} \right\rVert^2$ with $\alpha > 0$, and the network is globally synchronizable.
\end{proof}

\begin{theorem}\label{thm:QUAD_Gamma_semipositive_definite}    
Consider \eqref{eq:general_network} and assume that there exist 
$\B{P}, \B{Q}, \B{G} \in \BB{R}^{n \times n}$, $c \ge 0$, with $\B{P}, \B{G} > 0$, $\B{Q} = \B{Q}^- + \B{Q}'$, $\B{Q}^- < 0$, $\B{Q}' = (\B{Q}')\T$, such that
\begin{itemize}
    \item $\B{f}$ is QUAD($\B{P}$, $\B{Q}$);
    \item $\B{g}$ verifies Assumption \ref{ass:pseudo_QUAD_coupling} with $\B{P}$ and $\B{G}$;
    \item $\B{Q}'$ and $\B{G}$ are simultaneously diagonalisable;
    \item $\lambda_h(\B{G}) > 0$ if $\lambda_h(\B{Q}') > 0$, with $h = 1, \dots, n$.
\end{itemize}
Then, the network is globally synchronizable if
\begin{equation}
c \ge c^* \triangleq \begin{dcases}
\frac{1}{\lambda_{2} (\B{L})} \max_{h = 1, \dots, n} \frac{\lambda_h(\B{Q}')}{\lambda_h(\B{G})}, & \text{if } \exists h : \lambda_h(\B{Q}') > 0 \\
0, & \text{otherwise}
\end{dcases}.
\end{equation}    
\end{theorem}
\begin{proof}
The first part of the proof is identical to that of Theorem \ref{thm:QUAD_Gamma_positive_definite} until \eqref{eq:proof_branching_point_QUAD}, then we can write
\begin{equation}\label{eq:v_in_proof}
\begin{split}
v &\le \B{e}\T \left( \B{I}_N \otimes \B{Q} - c \B{L} \otimes \B{G} \right) \B{e} \\
&= \B{e}\T \left( \B{I}_N \otimes \B{Q}^- \right) \B{e} + \B{e}\T \left( \B{I}_N \otimes \B{Q}' \right) \B{e} - c v_\B{G},
\end{split}
\end{equation}
where $v_\B{G} \triangleq \B{e}\T \left( \B{L} \otimes \B{G} \right) \B{e}$.
Now, given that $\B{Q}'$ and $\B{G}$ are simultaneously diagonalisable, there exists an invertible matrix $\B{T} \in \BB{R}^{n \times n}$ such that $\B{Q}'=\B{T}^{-1} \B{\Delta}_{\B{Q}'} \B{T}$ and $\B{G}=\B{T}^{-1} \B{\Delta}_{\B{G}} \B{T}$, where $\B{\Delta}_{\B{Q}'}$ and $\B{\Delta}_{\B{G}}$ are diagonal matrices containing the real eigenvalues of $\B{Q}'$ and $\B{G}$, respectively (note that $\B{Q} = (\B{Q}')\T$ and $\B{G} = \B{G}\T$ imply that $\B{T}\T = \B{T}^{-1}$).
Let us also define the transformed synchronization errors $\B{y}_i \triangleq \B{T} \B{e}_i \in \BB{R}^n$ and their stack $\B{y} \triangleq \left( \B{I}_N \otimes \B{T} \right) \B{e} \in \BB{R}^{nN}$.
Therefore, we can rewrite $v_\B{G}$ as
\begin{equation*}
\begin{split}
v_\B{G} &= \B{e}\T \left( \B{L} \otimes \B{G} \right) \B{e}
= \B{e}\T \left[ \B{L} \otimes \left( \B{T}^{-1}\B{\Delta}_\B{G}\B{T} \right) \right] \B{e} \\
&= \B{e}\T \left[ 
\left( \B{L} \otimes \B{T}^{-1} \right)
\left( \B{I}_N \otimes (\B{\Delta}_\B{G}\B{T}) \right)
\right] \B{e} \\
&= \B{e}\T \left[ 
\left( \B{L} \otimes \B{T}^{-1} \right) 
\left( \B{I}_N \otimes \B{\Delta}_\B{G} \right)
\left( \B{I}_N \otimes \B{T} \right) 
\right] \B{e} \\
&= \B{e}\T \left[ 
\left( \B{I}_N \otimes \B{T}\T \right)
\left( \B{I}_N \otimes \B{T}\T \right)^{-1}
\left( \B{L} \otimes \B{T}^{-1} \right) 
\left( \B{I}_N \otimes \B{\Delta}_\B{G} \right) 
\right] \B{y} \\
&= \B{y}\T \left[ 
\left( \B{I}_N \otimes \B{T}\T \right)^{-1}
\left( \B{L} \otimes \B{T}^{-1} \right) 
\left( \B{I}_N \otimes \B{\Delta}_\B{G} \right) 
\right] \B{y} \\
&= \B{y}\T \left[ 
\left( \B{L} \otimes \left( \B{T} \B{T}^{-1} \right) \right) 
\left( \B{I}_N \otimes \B{\Delta}_\B{G} \right) 
\right] \B{y} 
= \B{y}\T \left( \B{L} \otimes \B{\Delta}_\B{G} \right) \B{y}. \\
\end{split}
\end{equation*}
Applying the same steps to $\B{e}\T \left( \B{I}_N \otimes \B{Q}' \right) \B{e}$, we rewrite \eqref{eq:v_in_proof} as
$
v \le \B{e}\T \left( \B{I}_N \otimes \B{Q}^- \right) \B{e} + \B{y}\T \left( \B{I}_N \otimes \B{\Delta}_{\B{Q}'} - c \B{L} \otimes \B{\Delta}_\B{G} \right) \B{y}.
$
Now, let us define $\B{y}_{*,h} \triangleq \begin{bmatrix} y_{1,h} & y_{2,h} & \cdots & y_{N,h} \end{bmatrix}\T \in \BB{R}^N$, with $h = 1, \dots, n$, as the vector of all the $h$-th components of the $N$ transformed synchronization errors $\B{y}_i$.
Since $\B{\Delta}_{\B{Q}'}$ and $\B{\Delta}_{\B{G}}$ are diagonal matrices, it is possible to write
\begin{equation*}
v \le \B{e}\T \left( \B{I}_N \otimes \B{Q}^- \right) \B{e} +
\sum\nolimits_{h=1}^{n} \B{y}_{*,h}\T \left[ \lambda_h(\B{Q}') \B{I}_N - c \lambda_h(\B{G}) \B{L} \right] \B{y}_{*,h},
\end{equation*}
and, using again Corollary 13.4.2 in \cite{godsil2013algebraic}, we have
\begin{equation*}
v \le \B{e}\T \left( \B{I}_N \otimes \B{Q}^- \right) \B{e} +
\sum\nolimits_{h=1}^{n} \left\lVert \B{y}_{*,h} \right \rVert^2 \left[ \lambda_h(\B{Q}') - c \lambda_h(\B{G}) \lambda_2(\B{L}) \right].
\end{equation*}
In order to have $v \le \B{e}\T \left( \B{I}_N \otimes \B{Q}^- \right) \B{e} < 0$, and thus prove synchronizability, it is required that $\lambda_h(\B{Q}') - c \lambda_h(\B{G}) \lambda_2(\B{L}) \le 0$, $h = 1, \dots, n$.
Note that if $\lambda_h(\B{Q}') \le 0$, then $\lambda_h(\B{G})$ can be null.
Differently, if $\lambda_h(\B{Q}') > 0$, then it is required that $\lambda_h(\B{G}) > 0$.
The value of $c^*$ stems trivially from the last consideration.
\end{proof}
As a handy simplification of Theorem \ref{thm:QUAD_Gamma_semipositive_definite}, we give the following corollary.
\begin{corollary}\label{cor:QUAD_Gamma_semipositive_definite}
Consider \eqref{eq:general_network} with linear diffusive coupling \eqref{eq:linear_diffusive_coupling} and assume that there exists $\B{Q} \in \BB{R}^{n \times n}$, with $\B{Q} = \B{Q}^- + \B{Q}'$, $\B{Q}^- < 0$, $\B{Q}' = \R{diag}\left( \begin{bmatrix} q_1 & \cdots & q_n \end{bmatrix} \right)$, such that
    \begin{itemize}
        \item $\B{f}$ is QUAD($\B{I}_n$, $\B{Q}$);
        \item  $\B{\Gamma}= \R{diag} \left( \begin{bmatrix} \gamma_1 & \cdots & \gamma_n \end{bmatrix} \right)$, with $\gamma_h \ge 0 \ \forall h = 1, \dots, n$, but $\gamma_h > 0$ if $q_h > 0$.
    \end{itemize}
    Then, the network is globally synchronizable if
    \begin{equation}
    c \ge c^* \triangleq \begin{dcases}
    \frac{1}{\lambda_{2} (\B{L})} \max_{h = 1, \dots, n} \frac{q_h}{\gamma_h}, & \text{if } \exists h : q_h > 0 \\
    0, & \text{otherwise}
    \end{dcases}.
    \end{equation}    
\end{corollary}
\begin{proof}
The proof is a direct consequence of Theorem \ref{thm:QUAD_Gamma_semipositive_definite}.
\end{proof}

\subsection*{Examples}
\begin{figure}[t]
    \centering
    \includegraphics[max width=\columnwidth]{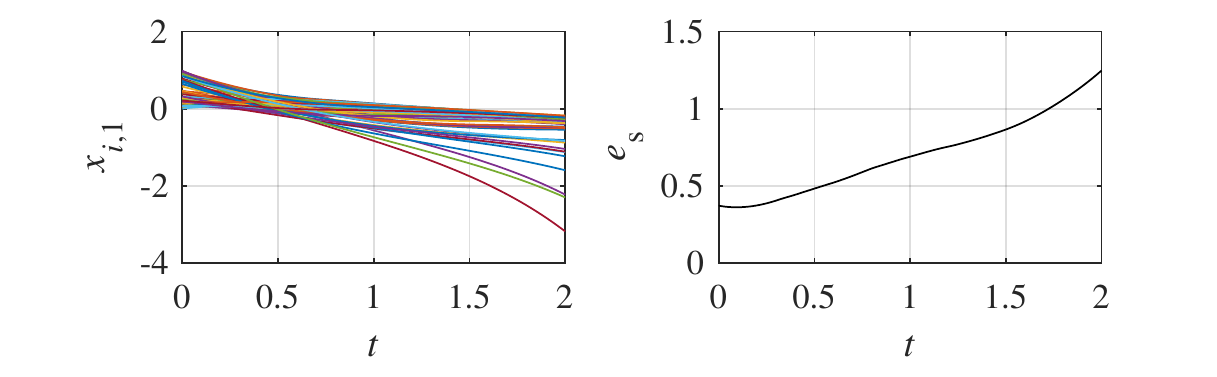} \\
    \includegraphics[max width=\columnwidth]{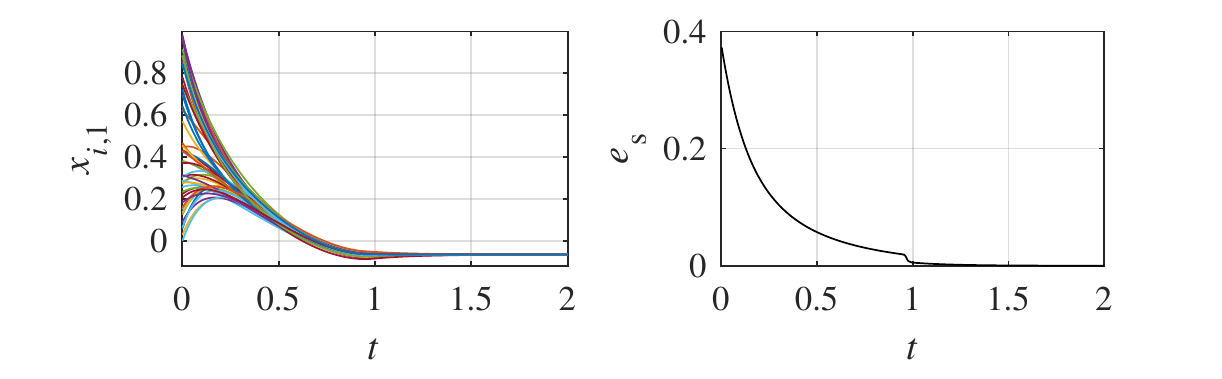}
    \caption{State dynamics and global synchronization error $e_\R{s}$ for coupled relay systems. Top panels: $c= 0.05$; bottom panels: $c = 0.25$.}
    \label{fig:quad}
\end{figure}

As an application of Theorem \ref{thm:QUAD_Gamma_positive_definite}, consider the classic relay system
$
\B{f}(\B{x}_i) = \left[ \begin{smallmatrix}
-1 & -1 \\
2 & 3
\end{smallmatrix} \right] \B{x}_i -
\left[ \begin{smallmatrix}
0 \\
2\R{sign}(x_{i,1} + x_{i,2})
\end{smallmatrix} \right].
$
Such system can either reach an equilibrium point in the set $\Omega = \{\B{x}_i : x_{i,1} = -x_{i,2}, \ x_{i,2} \in [-2,2] \}$ or diverge, and is QUAD with $\B{P} = \B{I}_n$ and $\B{Q} = 3.06 \B{I}_n$.
We coupled $N = 50$ of these relays through an Erd\"os-Rényi random graph with probability $p=0.5$ \cite{erdos1959random}, resulting in a topology with $\lambda_2(\B{L}) = 14.80$; in addition, we considered a linear diffusive coupling \eqref{eq:linear_diffusive_coupling} with $\B{\Gamma} = \B{I}_n$.
The critical value of the coupling gain computed using Theorem \ref{thm:QUAD_Gamma_positive_definite} is $c^* = \left\lVert \B{Q} \right\rVert / \lambda_2(\B{L}) = 0.21$. 
Figure \ref{fig:quad} shows the absence and the emergence of synchronization in the cases $c = 0.05 < c^*$ and $c = 0.25 > c^*$.

\begin{figure}[t]
    \centering
    \includegraphics[max width=\columnwidth]{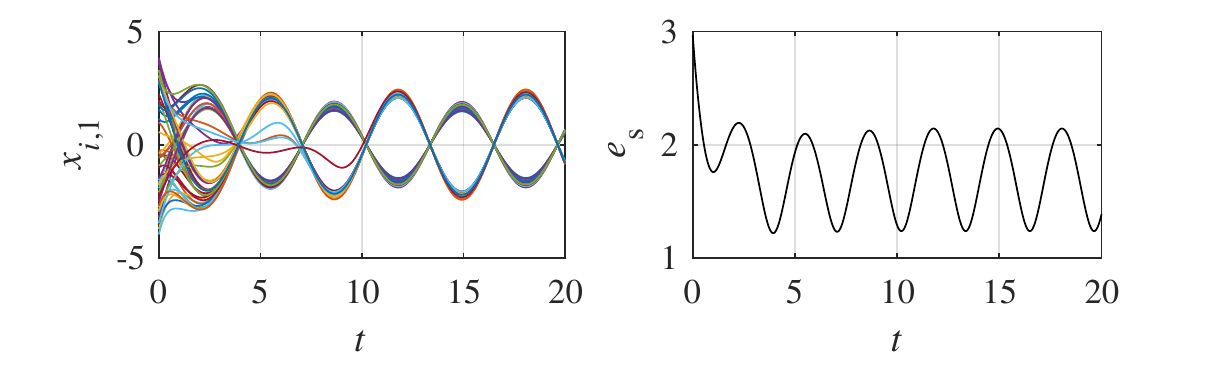} \\
    \includegraphics[max width=\columnwidth]{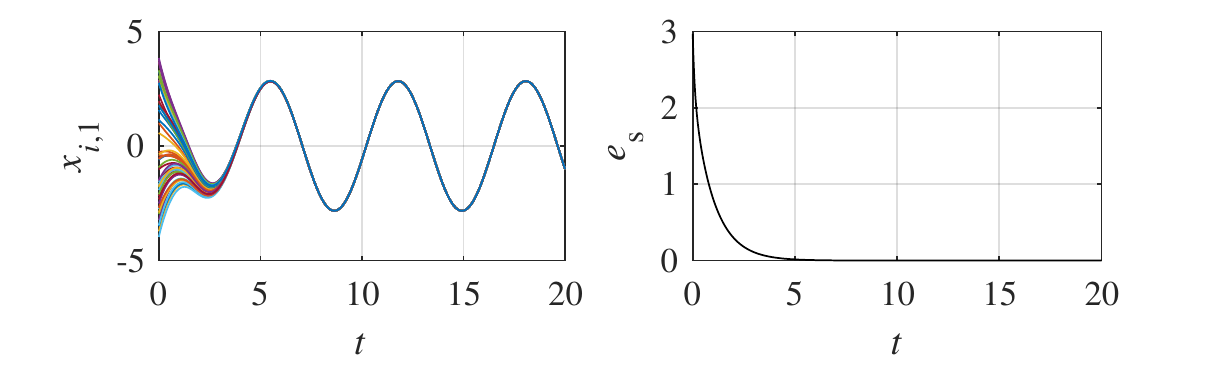}
    \caption{State dynamics and global synchronization error $e_\R{s}$ for coupled oscillating systems. Top panels: $c= 0.02$; bottom panels: $c = 0.28$.}
    \label{fig:oscillator}
\end{figure}

Then, to illustrate Theorem \ref{thm:QUAD_Gamma_semipositive_definite} and Corollary \ref{cor:QUAD_Gamma_semipositive_definite}, consider the following PWS oscillator as a representative example:
$
\B{f}(\B{x}_i; t) = \left[ \begin{smallmatrix}
- x_{i,1} + 2 x_{i,2} \R{sin}(t) \\
f_2(x_{i,2})
\end{smallmatrix} \right]
$,
where 
$
f_2(x_{i,2}) = \left\{
\begin{smallmatrix}
-x_{i,2} - 2, \hfill & x_{i,2} \le -1   \hfill \\
x_{i,2},      \hfill & -1 < x_{i,2} < 1 \hfill \\
-x_{i,2} + 2, \hfill & x_{i,2} \ge 1    \hfill \\
\end{smallmatrix}\right.
$.
This is a cascaded system, as $\dot{x}_{i,2}$ depends only on $x_{i,2}$. 
Moreover the state variable $x_{i,2}$ has two stable equilibria in $-2$ and $+2$; $x_{i,1}$ displays a sinusoidal behaviour, whose amplitude and phase are dependant on $x_{i,2}$.
Notice that $\B{f}$ is continuous but not differentiable, and QUAD with $\B{P} = \B{I}_n$ and
$
\B{Q} = \left[ \begin{smallmatrix}
-1 & 2 \\
0 & 1
\end{smallmatrix} \right]
$;
then we take
$
\B{Q}^- = \left[ \begin{smallmatrix}
-1 & 2 \\
0 & -3
\end{smallmatrix} \right]
$
and
$
\B{Q}' = \left[ \begin{smallmatrix}
0 & 0 \\
0 & 4
\end{smallmatrix} \right]
$.
As in the previous example, we deploy a random network with $N=50$ nodes, and again $\lambda_2(\B{L}) = 14.80$, but this time
$
\B{\Gamma} = \left[ \begin{smallmatrix}
0 & 0 \\
0 & 1
\end{smallmatrix} \right]
$ (note that $\B{\Gamma} \not> 0$).
Applying Theorem \ref{thm:QUAD_Gamma_semipositive_definite} with $\B{T} = \B{I}_n$ we get $c^* = \lambda_{2} (\B{Q}') / [ \lambda_{2} (\B{L}) \lambda_{2} (\B{\Gamma}) ] = 4 / [ 14.80 \cdot 1] = 0.27$.
Figure \ref{fig:oscillator} shows the results of two simulations, with $c = 0.02 < c^*$, and $c = 0.28 > c^*$; only the latter case displays synchronization.

\subsection{Extension to non-QUAD systems}
When $\B{f}$ is not QUAD, but satisfies the milder Assumption \ref{ass:pseudo_QUAD} instead, a discontinuous coupling action can be added to a standard linear diffusive coupling in order to enable global synchronizability.
Here, we study the following pair of coupled agents as a paradigm to understand emerging properties in larger networks:
\begin{equation}\label{eq:two_nodes_pws}
\dot{\B{x}}_i = 
\B{f}(\B{x}_i;t) + c \B{\Gamma}(\B{x}_j - \B{x}_i) + c_\R{d} \B{\Gamma}_\R{d} \R{sign}(\B{x}_j - \B{x}_i),
\end{equation}
$i, j = 1, 2, \ i \ne j$. In particular, in Theorems \ref{thm:pseudo_QUAD_Gamma_positive_definite} and \ref{thm:pseudo_QUAD_Gamma_semipositive_definite} we give conditions for global synchronizability; the former theorem is meant to be used when $\B{\Gamma} > 0$, whereas the latter can be utilised when no assumptions on the definiteness of $\B{\Gamma}$ can be made.

\begin{theorem}\label{thm:pseudo_QUAD_Gamma_positive_definite}
    Consider \eqref{eq:two_nodes_pws} and assume that there exist $\B{P}, \B{Q} \in \BB{R}^{n \times n}$, $\B{m} \in \BB{R}^n$, with $\B{P} > 0$, $\B{m} \ne \B{0}$, such that
    \begin{itemize}
        \item $\B{f}$ verifies Assumption \ref{ass:pseudo_QUAD};
        \item $\R{sym}(\B{P} \B{\Gamma}) > 0$;
        \item $\B{P}\B{\Gamma}_\R{d} = \R{diag}(\B{\gamma}_\R{d})$, with $\B{\gamma}_\R{d} = \begin{bmatrix} \gamma_{\R{d},1} & \cdots & \gamma_{\R{d},n} \end{bmatrix}\T \in \BB{R}^n$, and $\gamma_{\R{d},h} \ge 0 \ \forall h = 1, \dots, n$, but $\gamma_{\R{d},h} > 0$ if $m_h > 0$.
    \end{itemize}
    Then, \eqref{eq:two_nodes_pws} is globally synchronizable if 
    \begin{equation}
    c > c^* \triangleq \frac{\left\lVert \B{Q} \right \rVert}{2 \lambda_{\R{min}} \left[ \R{sym} (\B{P} \B{\Gamma}) \right] }, \quad
    c_\R{d} \ge c^*_\R{d} \triangleq \frac{1}{2} \max_{h= 1, \dots, n} \frac{m_h}{\gamma_{\R{d},h}}.
    \end{equation}
\end{theorem}
\begin{proof}
Consider the candidate common Lyapunov function $V(\B{e}) = \frac{1}{2} \B{e}\T \B{P} \B{e}$, where $\B{e} \triangleq \B{x}_1 - \B{x}_2$.
Then
\begin{equation*}
\begin{split}
\dot{V} &= \B{e}\T\B{P}\dot{\B{e}} = 
\B{e}\T \B{P} \left[ \B{f} (\B{x}_1;t) - \B{f}(\B{x}_2;t) \right] \\
&\phantom{=\ }+ \B{e}\T \B{P}\left[ 2 c \B{\Gamma} (\B{x}_2 - \B{x}_1) + 2 c_\R{d} \B{\Gamma}_\R{d} \R{sign} (\B{x}_2 - \B{x}_1) \right] \\
&= \B{e}\T \B{P} [\B{f} (\B{x}_1;t) - \B{f}(\B{x}_2;t) ] + \B{e}\T \B{P} \left[ - 2 c \B{\Gamma} \B{e} - 2 c_\R{d} \B{\Gamma}_\R{d} \R{sign} (\B{e}) \right].
\end{split}
\end{equation*}
Using Assumption \ref{ass:pseudo_QUAD} and the fact that $\B{P}\B{\Gamma}_\R{d} = \R{diag}(\B{\gamma}_\R{d})$, we have
\begin{equation*}
\begin{split}
\dot{V} &\le \B{e}\T \B{Q} \B{e} + \B{m}\T \lvert \B{e} \rvert - 2 c \B{e}\T \B{P} \B{\Gamma} \B{e} - 2 c_\R{d} \B{\gamma}_\R{d}\T \lvert \B{e} \rvert \\
&= \B{e}\T (\B{Q} - 2 c \B{P} \B{\Gamma})\B{e} + (\B{m} - 2 c_\R{d} \B{\gamma}_\R{d})\T \lvert \B{e} \rvert \\
&\le \left\lVert \B{e} \right\rVert^2 \left[ \lVert \B{Q} \rVert - 2c \lambda_{\R{min}} (\R{sym}(\B{P} \B{\Gamma})) \right] \\
&\phantom{={}\ } + \sum\nolimits_{h=1}^{n}\left[(m_h - 2 c_\R{d} \gamma_{\R{d},h}) \lvert e_h \rvert \right].
\end{split}
\end{equation*}
Therefore, it is immediate to verify that if $c > c^*$ and $c \ge c_\R{d}^*$, then $\dot{V}(\B{e}) < - \alpha \left\lVert \B{e} \right\rVert^2$, with $\alpha > 0$, and the pair of agents is globally synchronizable.
\end{proof}

\begin{theorem}\label{thm:pseudo_QUAD_Gamma_semipositive_definite}
Consider \eqref{eq:two_nodes_pws} and assume that there exist $\B{P}, \B{Q} \in \BB{R}^{n \times n}$, $\B{m} \in \BB{R}^n$, with $\B{P} > 0$, $\B{Q} = \B{Q}^- + \B{Q}'$, $\B{Q}^- < 0$, $\B{Q}' = (\B{Q}')\T$, $\B{m} \ne \B{0}$, such that
\begin{itemize}
    \item $\B{f}$ verifies Assumption \ref{ass:pseudo_QUAD};
    \item $\B{Q}'$ and $\B{G} \triangleq \B{P}\B{\Gamma}$ are simultaneously diagonalisable;
    \item $\lambda_h(\B{G}) > 0$ if $\lambda_h(\B{Q}') > 0$, for $h = 1, \dots, n$.
    \item $\B{P}\B{\Gamma}_\R{d} = \R{diag}(\B{\gamma}_\R{d})$, with $\B{\gamma}_\R{d} = \begin{bmatrix} \gamma_{\R{d},1} & \cdots & \gamma_{\R{d},n} \end{bmatrix}\T \in \BB{R}^n$, and $\gamma_{\R{d},h} \ge 0 \ \forall h = 1, \dots, n$, but $\gamma_{\R{d},h} > 0$ if $m_h > 0$.
\end{itemize}
Then, \eqref{eq:two_nodes_pws} is globally synchronizable if
\begin{equation}
\begin{aligned}
c &> c^* \triangleq \begin{dcases}
\frac{1}{2} \max_{h = 1, \dots, n} \frac{\lambda_h(\B{Q}')}{\lambda_h(\B{G})}, & \text{if } \exists h : \lambda_h(\B{Q}') > 0 \\
0, & \text{otherwise}
\end{dcases}, \\
c_\R{d} &\ge c^*_\R{d} \triangleq \frac{1}{2} \max_{h = 1, \dots, n} \frac{m_h}{\gamma_{\R{d},h}}.
\end{aligned}
\end{equation}
\end{theorem}
\begin{proof}
The proof is obtained simply by combining those of Theorems \ref{thm:QUAD_Gamma_semipositive_definite} and \ref{thm:pseudo_QUAD_Gamma_positive_definite}, and thus omitted for brevity.    
\end{proof}

\subsection*{Example}

\begin{figure}[t]
    \centering
    \includegraphics[max width=\columnwidth]{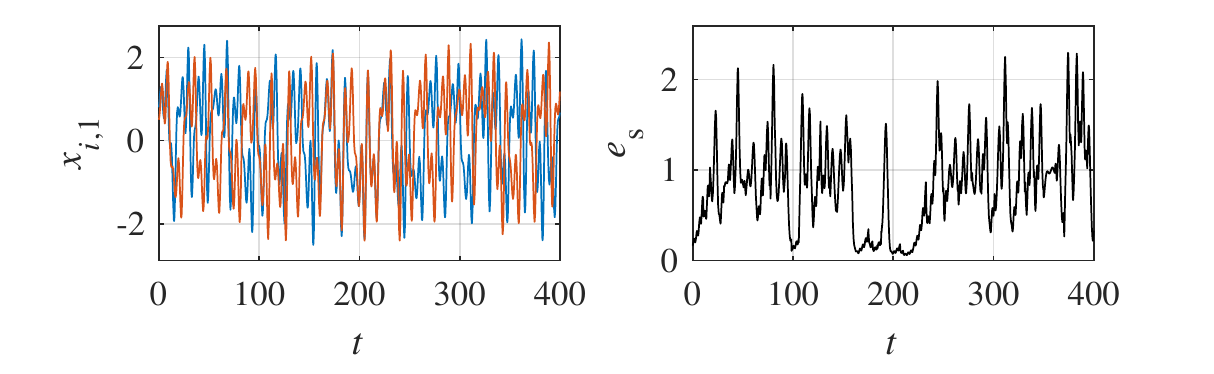}\\
    \includegraphics[max width=\columnwidth]{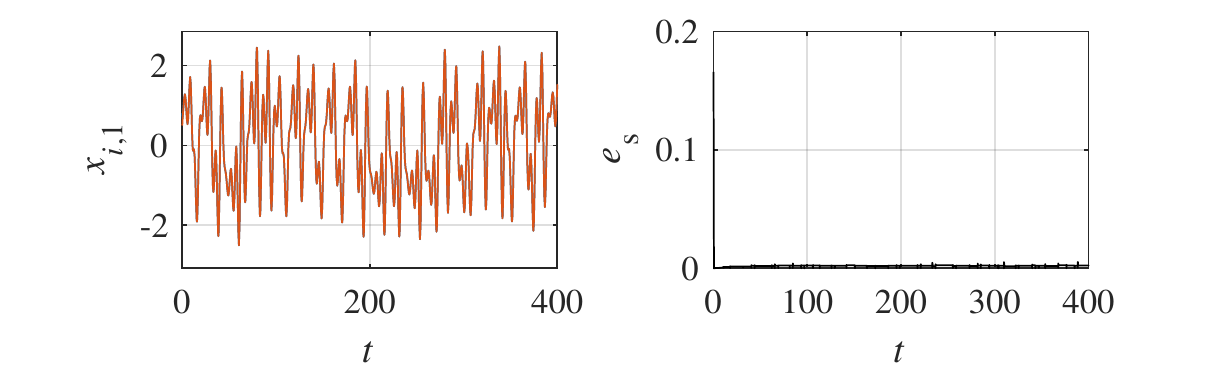}
    \caption{State dynamics (blue is $x_{1,1}$, orange is $x_{2,1}$) and global synchronization error $e_\R{s}$ for coupled Sprott circuits; Top panels: $c = 0.002 c^*$, $c_\R{d} = 0.002 c_\R{d}^*$; bottom panels: $c = 1.002 c^*$, $c_\R{d} = 1.002 c_\R{d}^*$.}
    \label{fig:sprott_N2}
\end{figure}

To illustrate Theorem \ref{thm:pseudo_QUAD_Gamma_positive_definite}, we consider a network of two chaotic Sprott circuits \cite{sprott2000anew}, whose dynamics is described by
\begin{equation}\label{eq:sprott_circuit}
\B{f}(\B{x}_i) = \left[\begin{smallmatrix}
0 & 1 & 0 \\
0 & 0 & 1 \\
-1 & -1 & -0.5 
\end{smallmatrix}\right]
\B{x}_i
+ \left[ \begin{smallmatrix}
0 \\
0  \\
\R{sign}(x_{i,1})
\end{smallmatrix} \right],
\end{equation}
coupled through the matrices $\B{\Gamma} = \B{\Gamma}_\R{d} = \B{I}_n$.
In this scenario, $\B{P} = \B{I}_n$, $\left\lVert \B{Q} \right\rVert = 1.70$, $\lambda_{\R{min}} \left( \R{sym} (\B{P} \B{\Gamma}) \right) = 1$, and $\B{m} = \begin{bmatrix} 2 & 0 & 0 \end{bmatrix}\T$; hence, $c^* = 0.85$ and $c_\R{d}^* = 1$; as initial condition we selected $\B{x}_1(0) = \begin{bmatrix} 0.8 & 0.2 & 0.2 \end{bmatrix}\T$, $\B{x}_2(0) = \begin{bmatrix} 0.5 & 0.1 & 0.1 \end{bmatrix}\T$.
Figure \ref{fig:sprott_N2} depicts the results of two simulations: in the former $c = 0.002 c^*$ and $c_\R{d} = 0.002 c_\R{d}^*$, whereas in the latter $c = 1.002 c^*$ and $c_\R{d} = 1.002 c_\R{d}^*$.
Synchronization is achieved only in the second case, where $ c > c^*$ and $c_\R{d} > c_\R{d}^*$.

\section{Multiplex networks of $N$ generic PWS systems}
\label{sec:multiplex_networks}

\subsection{A switched multiplex approach}

When $N \ge 3$ agents are present in the network and the QUAD assumption is not fulfilled by the vector field $\B{f}$, we propose to extend \eqref{eq:two_nodes_pws} using a \emph{multiplex} network approach, inspired by the strategy used in \cite{lombana2016multiplex}, to enforce consensuability in networks of smooth systems.
Specifically, we consider networks in which the coupling between nodes consists of two layers: (i) a diffusive coupling layer with topology described by the matrix $\B{L}$, and (ii) a discontinuous coupling layer, possibly characterized by a different topology, encoded by the Laplacian $\B{L}_\R{d}$.
Namely, the overall network dynamics becomes
\begin{equation}\label{eq:network_disc_coup}
\dot{\B{x}}_i = 
\B{f}(\B{x}_i;t)
- c \sum_{j=1}^{N} L_{ij} \B{\Gamma} (\B{x}_j - \B{x}_i)
- c_\R{d} \sum_{j=1}^{N} L_{ij}^\R{d} \B{\Gamma}_\R{d} \R{sign} (\B{x}_j - \B{x}_i), \end{equation}
with $i = 1, \dots, N$, and $L_{ij}^\R{d}$ being the element $(i,j)$ of the symmetric Laplacian matrix $\B{L}_\R{d}$ associated to the graph $\C{G}_\R{d}$ relative to the discontinuous coupling.
A complete proof of convergence of this multiplex approach is beyond the scope of this paper and will be presented elsewhere.
Next, we proceed with an exhaustive numerical analysis to illustrate how the choice of the structure of the coupling layers can affect the stability and synchronizability of the network.

\subsection{Numerical study}

\begin{figure}[t]
    \sidesubfloat[]{\includegraphics[max width=\columnwidth]{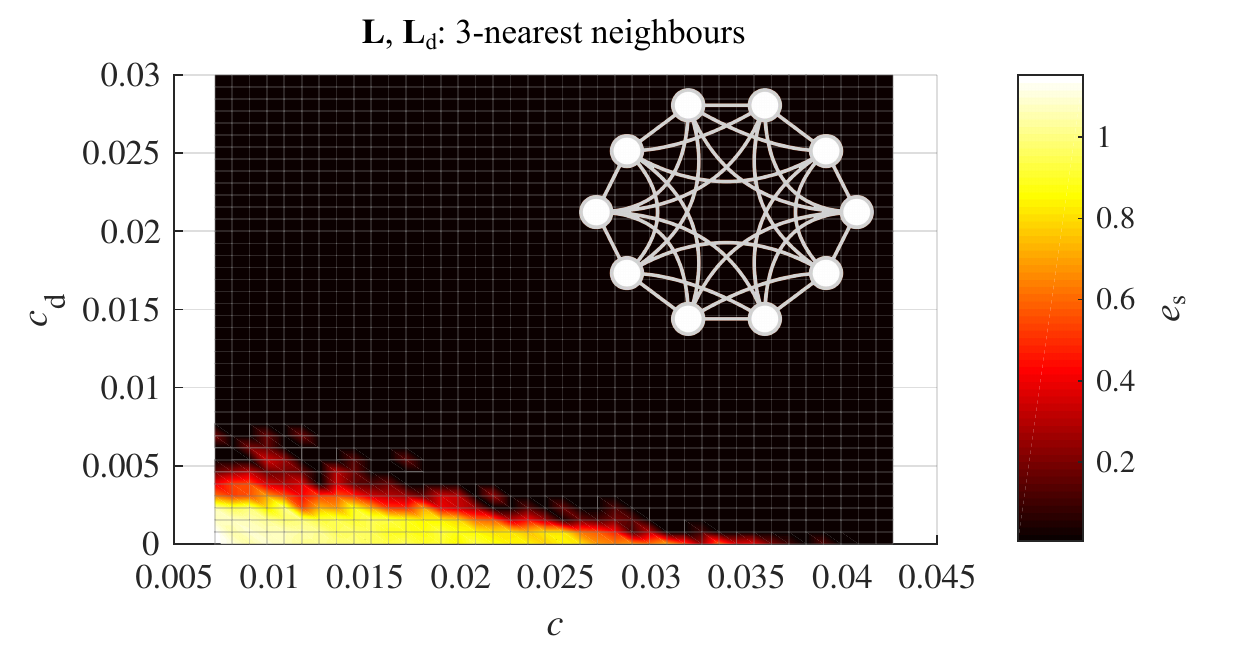}
        \label{fig:sprott_a}}\\%
    \sidesubfloat[]{\includegraphics[max width=\columnwidth]{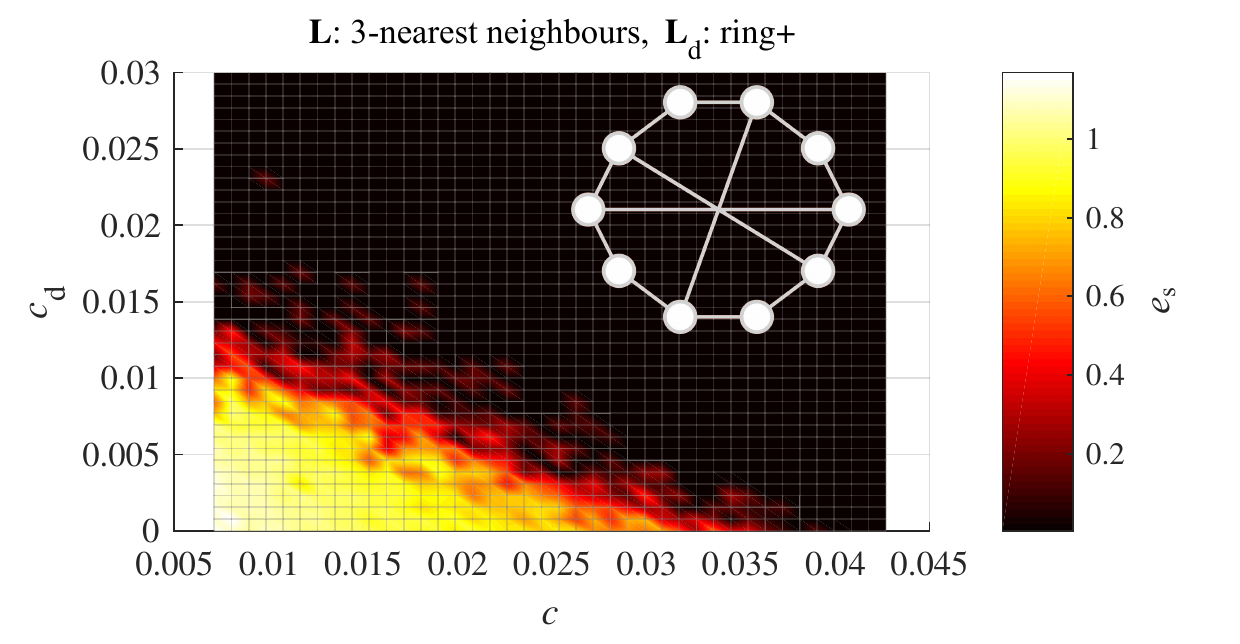}
        \label{fig:sprott_b}}\\%
    \sidesubfloat[]{\includegraphics[max width=\columnwidth]{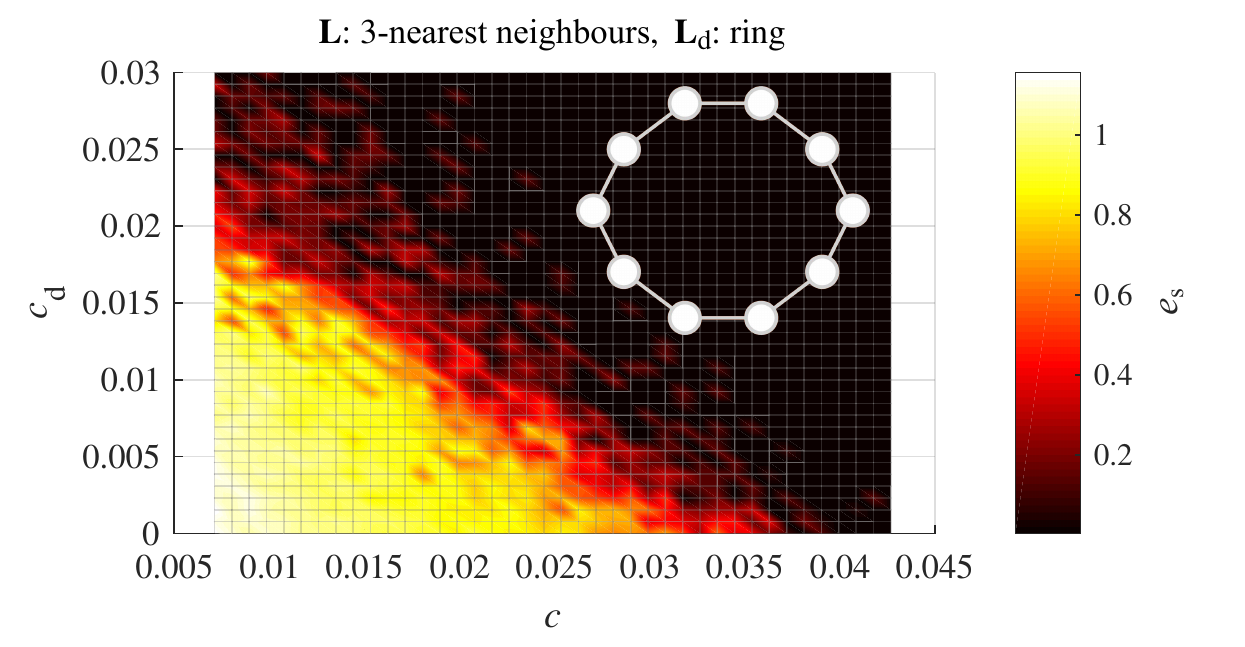}
        \label{fig:sprott_c}}
    \caption{Characterization of synchronizability in a network of Sprott circuits. Five random initial conditions were used, with $\B{x}_i(0) \in [ [0,1] \ [0, 0.5] \ [0, 0.5] ]\T$, and, for each combination of $c$ and $c_\R{d}$, $e_\R{s}$ is taken as the average of the five simulations. The diffusive layer is always associated to a 3-nearest neighbours; differently, the discontinuous coupling layer varies in each figure.}
    \label{fig:sprott}
\end{figure}

To provide a proof of the enhanced synchronizability provided by the discontinuous coupling in \eqref{eq:network_disc_coup}, we consider a network of $N = 10$ identical Sprott circuits \eqref{eq:sprott_circuit}.
In the network, $\B{\Gamma} = \B{\Gamma}_\R{d} = \B{I}_n$, and the nodes are diffusively coupled via a graph with Laplacian matrix $\B{L}$, associated to a 3-nearest neighbours topology.
Differently, $\B{L}_\R{d}$ is associated to 3 possible topologies, as portrayed in Figure \ref{fig:sprott}, which displays the steady state value of the global synchronization error $e_\R{s}$ (defined in Section \ref{sec:network_model}) for each different combination of the coupling layer structures. 
Initial conditions were selected randomly (with a uniform distribution), in the range of chaoticity of the Sprott circuit.
We notice that the stability region depends on the relative choice of the structures of the two coupling layers.
Obviously, the worst case is when the structure of the discontinuous layer is the sparsest (see Figure \ref{fig:sprott_c}).
Surprisingly, to enhance stability it is sufficient to add a few long range links to the discontinuous coupling layer (see Figure \ref{fig:sprott_b}); the largest stability region being observed when the discontinuous coupling layer shares the same links as the underlying diffusive one.

\subsection{A further example}

\begin{figure}[t]
    \centering
    \includegraphics[max width=\columnwidth]{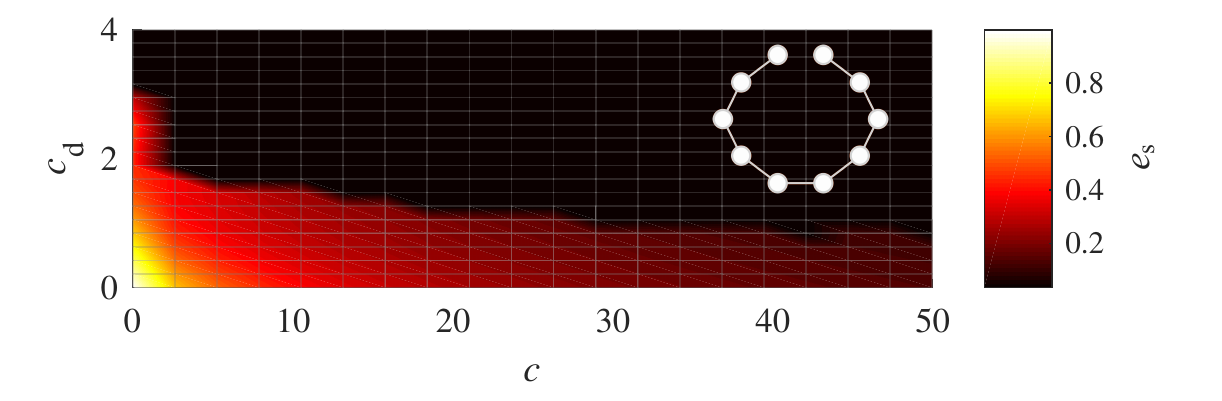}
        \label{fig:bistables_a}
    \caption{Characterization of synchronizability in a network of bistable systems.}
    \label{fig:bistables}
\end{figure}

To further illustrate the beneficial effect of the discontinuous layer, we consider a network of $N = 10$ PWS bistable systems, used to model energy harvesters \cite{cohen2014onthe} or simplified climatic models \cite{leifeld2015persistence}, and described by
$
\B{f}(\B{x}_i) = \left[\begin{smallmatrix}
0 & 1 \\
-1 & -1
\end{smallmatrix}\right]
\B{x}_i
+ \left[\begin{smallmatrix}
0  \\
\R{sign}(x_{i,1})
\end{smallmatrix}\right]
$.
The system has two coexisting stable equilibria in $\begin{bmatrix} 1 & 0 \end{bmatrix}\T$ and $\begin{bmatrix} -1 & 0 \end{bmatrix}\T$.
The agents are coupled over a path graph, with $\B{L} = \B{L}_\R{d}$ and $\B{\Gamma} = \B{\Gamma}_\R{d} = \B{I}_n$.
We consider the particularly challenging case where five nodes are started at one of the equilibria, while the other five are at the other.
In this case, as shown in Figure \ref{fig:bistables}, the diffusive coupling layer alone is unable to synchronize the network for any value of $c$, when the discontinuous coupling layer is disconnected ($c_\R{d} = 0$).
Synchronization in this case is only achieved when both coupling layers are present.

\section{Conclusion}
\label{sec:conclusion}

We have discussed the problem of complete spontaneous synchronizability in networks of piecewise-smooth systems.
Specifically, we started by providing sufficient conditions for ensembles of QUAD PWS systems, applicable to problems with a large variety of coupling laws, including linear diffusive coupling with indefinite inner coupling matrix.
Next, we found that, for two coupled agents, when their dynamics is not QUAD, a discontinuous coupling (added to a linear diffusive coupling) can be used to enforce synchronizability.
Motivated by this finding, we then extended the study numerically to larger networks of $N$ nodes, allowing for a multiplex structure, which means the presence of different topologies for different coupling actions.
Targeted and extensive numerical analyses illustrated the effectiveness of the multiplex approach: the discontinuous layer, even when associated to a sparse topology, makes complete synchronization feasible also when a linear coupling protocol alone cannot.

\bibliographystyle{IEEEtran}  
\bibliography{references_cdc_2018}

\end{document}